\documentclass[%
 reprint,
superscriptaddress,
 amsmath,amssymb,
 aps,
]{revtex4-2}

\usepackage{graphicx}
\usepackage{dcolumn}
\usepackage{bm}

\usepackage{xcolor}
\usepackage{cancel}

\begin{document}

\preprint{APS/123-QED}

\title{First measurement of the neutron-emission probability with a surrogate reaction in inverse kinematics at a heavy-ion storage ring\\}

\author{M. Sguazzin}
\thanks{Present address: IJCLab, 91405 Orsay, France}
\affiliation{Université de Bordeaux, CNRS, LP2I Bordeaux, 33170 Gradignan, France}
\author{B. Jurado}
\thanks{Corresponding author, jurado@cenbg.in2p3.fr}
\affiliation{Université de Bordeaux, CNRS, LP2I Bordeaux, 33170 Gradignan, France}
\author{J. Pibernat}
\affiliation{Université de Bordeaux, CNRS, LP2I Bordeaux, 33170 Gradignan, France}
\author{J. A. Swartz}
\thanks{Present address: FRIB, MSU, Michigan 48824, USA}
\affiliation{Université de Bordeaux, CNRS, LP2I Bordeaux, 33170 Gradignan, France}
\author{M. Grieser}
\affiliation{Max-Planck-Institut für Kernphysik, 69117 Heidelberg, Germany}
\author{J. Glorius}
\affiliation{GSI Helmholtzzentrum für Schwerionenforschung, 64291 Darmstadt, Germany}
\author{\mbox{Yu. A. Litvinov}}
\affiliation{GSI Helmholtzzentrum für Schwerionenforschung, 64291 Darmstadt, Germany}
\author{\mbox{J. Adamczewski-Musch}}
\affiliation{GSI Helmholtzzentrum für Schwerionenforschung, 64291 Darmstadt, Germany}
\author{P. Alfaurt}
\affiliation{Université de Bordeaux, CNRS, LP2I Bordeaux, 33170 Gradignan, France}
\author{P. Ascher}
\affiliation{Université de Bordeaux, CNRS, LP2I Bordeaux, 33170 Gradignan, France}
\author{L. Audouin}
\affiliation{Université Paris-Saclay, CNRS, IJCLab, 91405 Orsay, France}
\author{C. Berthelot} 
\affiliation{Université de Bordeaux, CNRS, LP2I Bordeaux, 33170 Gradignan, France}
\author{B. Blank}
\affiliation{Université de Bordeaux, CNRS, LP2I Bordeaux, 33170 Gradignan, France}
\author{\mbox{K. Blaum}} 
\affiliation{Max-Planck-Institut für Kernphysik, 69117 Heidelberg, Germany}
\author{\mbox{B. Br\"uckner}}
\affiliation{Goethe University of Frankfurt, 60438 Frankfurt, Germany}
\author{\mbox{S. Dellmann}}
\affiliation{Goethe University of Frankfurt, 60438 Frankfurt, Germany}
\author{I. Dillmann}
\affiliation{TRIUMF, Vancouver, British Columbia, V6T 2A3, Canada}
\affiliation{Department of Physics and Astronomy, University of Victoria, Victoria, BC, V8P 5C2, Canada}
\author{C. Domingo-Pardo} 
\affiliation{IFIC, CSIC-Universidad de Valencia, 46980 Valencia, Spain}
\author{M. Dupuis} 
\affiliation{CEA, DAM, DIF, 91297 Arpajon, France}
\affiliation{Université Paris-Saclay, CEA, LMCE, 91680 Bruyères-Le-Châtel, France}
\author{P. Erbacher}
\affiliation{Goethe University of Frankfurt, 60438 Frankfurt, Germany}
\author{M. Flayol} 
\affiliation{Université de Bordeaux, CNRS, LP2I Bordeaux, 33170 Gradignan, France}
\author{O. Forstner}
\affiliation{GSI Helmholtzzentrum für Schwerionenforschung, 64291 Darmstadt, Germany}
\author{\mbox{D. Freire-Fernández}}
\affiliation{Max-Planck-Institut für Kernphysik, 69117 Heidelberg, Germany}
\affiliation{Ruprecht-Karls-Universität Heidelberg, 69117 Heidelberg, Germany}
\author{M. Gerbaux}
\affiliation{Université de Bordeaux, CNRS, LP2I Bordeaux, 33170 Gradignan, France}
\author{J. Giovinazzo} 
\affiliation{Université de Bordeaux, CNRS, LP2I Bordeaux, 33170 Gradignan, France}
\author{S. Grévy}
\affiliation{Université de Bordeaux, CNRS, LP2I Bordeaux, 33170 Gradignan, France}
\author{C. J. Griffin}
\affiliation{TRIUMF, Vancouver, British Columbia, V6T 2A3, Canada}
\author{A. Gumberidze}
\affiliation{GSI Helmholtzzentrum für Schwerionenforschung, 64291 Darmstadt, Germany}
\author{S. Heil} 
\affiliation{Goethe University of Frankfurt, 60438 Frankfurt, Germany}
\author{\mbox{A. Heinz}}
\affiliation{Chalmers University of Technology, 41296 Gothenburg, Sweden}
\author{\mbox{R. Hess}}
\affiliation{GSI Helmholtzzentrum für Schwerionenforschung, 64291 Darmstadt, Germany}
\author{\mbox{D. Kurtulgil}} 
\affiliation{Goethe University of Frankfurt, 60438 Frankfurt, Germany}
\author{\mbox{N. Kurz}}
\affiliation{GSI Helmholtzzentrum für Schwerionenforschung, 64291 Darmstadt, Germany}
\author{G. Leckenby}
\affiliation{TRIUMF, Vancouver, British Columbia, V6T 2A3, Canada}
\affiliation{Department of Physics and Astronomy, University of British Columbia, Vancouver, BC, V6T 1Z1, Canada}
\author{S. Litvinov} 
\affiliation{GSI Helmholtzzentrum für Schwerionenforschung, 64291 Darmstadt, Germany}
\author{B. Lorentz}
\affiliation{GSI Helmholtzzentrum für Schwerionenforschung, 64291 Darmstadt, Germany}
\author{V. Méot} 
\affiliation{CEA, DAM, DIF, 91297 Arpajon, France}
\affiliation{Université Paris-Saclay, CEA, LMCE, 91680 Bruyères-Le-Châtel, France}
\author{\mbox{J. Michaud}}
\thanks{Present address: IJCLab, 91405 Orsay, France}
\affiliation{Université de Bordeaux, CNRS, LP2I Bordeaux, 33170 Gradignan, France}
\author{\mbox{S. Pérard}}
\affiliation{Université de Bordeaux, CNRS, LP2I Bordeaux, 33170 Gradignan, France}
\author{\mbox{N. Petridis}}
\affiliation{GSI Helmholtzzentrum für Schwerionenforschung, 64291 Darmstadt, Germany}
\author{U. Popp}
\affiliation{GSI Helmholtzzentrum für Schwerionenforschung, 64291 Darmstadt, Germany}
\author{D. Ramos}
\affiliation{GANIL, CRNS/IN2P3-CEA/DRF, 14000 Caen, France}
\author{R. Reifarth} 
\affiliation{Goethe University of Frankfurt, 60438 Frankfurt, Germany}
\affiliation{Los Alamos National Laboratory, Los Alamos, NM, 87544, USA}
\author{M. Roche}
\affiliation{Université de Bordeaux, CNRS, LP2I Bordeaux, 33170 Gradignan, France}
\author{M.S. Sanjari} 
\affiliation{GSI Helmholtzzentrum für Schwerionenforschung, 64291 Darmstadt, Germany}
\affiliation{Aachen University of Applied Sciences, Aachen, Germany}
\author{\mbox{R.S. Sidhu}} 
\affiliation{School of Physics and Astronomy, University of Edinburgh, EH9 3FD Edinburgh, United Kingdom}
\affiliation{GSI Helmholtzzentrum für Schwerionenforschung, 64291 Darmstadt, Germany}
\affiliation{Max-Planck-Institut für Kernphysik, 69117 Heidelberg, Germany}
\author{\mbox{U. Spillmann}} 
\affiliation{GSI Helmholtzzentrum für Schwerionenforschung, 64291 Darmstadt, Germany}
\author{M. Steck}  
\affiliation{GSI Helmholtzzentrum für Schwerionenforschung, 64291 Darmstadt, Germany}
\author{Th. Stöhlker} 
\affiliation{GSI Helmholtzzentrum für Schwerionenforschung, 64291 Darmstadt, Germany}
\author{B. Thomas}
\affiliation{Université de Bordeaux, CNRS, LP2I Bordeaux, 33170 Gradignan, France}
\author{L. Thulliez}
\affiliation{IRFU, CEA, Université Paris-Saclay, 91191 Gif-sur-Yvette, France}
\author{M. Versteegen}
\affiliation{Université de Bordeaux, CNRS, LP2I Bordeaux, 33170 Gradignan, France}
\author{B. Włoch}
\affiliation{Université de Bordeaux, CNRS, LP2I Bordeaux, 33170 Gradignan, France}


\date{\today}

\begin{abstract}
Neutron-induced reaction cross sections of short-lived nuclei are imperative to understand the origin of heavy elements in stellar nucleosynthesis and for societal applications, but their measurement is extremely complicated due to the radioactivity of the targets involved. One way of overcoming this issue is to combine surrogate reactions with the unique possibilities offered by heavy-ion storage rings. In this work, we describe the first surrogate-reaction experiment in inverse kinematics, which we successfully conducted at the Experimental Storage Ring (ESR) of the GSI/FAIR facility, using the $^{208}$Pb(p,p’) reaction as a surrogate for neutron capture on $^{207}$Pb. Thanks to the outstanding detection efficiencies possible at the ESR, we were able to measure for the first time the neutron-emission probability as a function of the excitation energy of $^{208}$Pb. We have used this probability to select different descriptions of the $\gamma$-ray strength function and nuclear level density, and provide reliable results for the neutron-induced radiative capture cross section of $^{207}$Pb at energies for which no experimental data exist.
\end{abstract}

\maketitle


Knowledge of neutron-induced reaction cross sections of short-lived nuclei is pivotal to our understanding of the synthesis of elements via the astrophysical slow (s) and rapid (r) neutron capture processes, about which there are still many uncertainties and open questions \cite{Arn20}. It is also of interest for applications such as nuclear waste management and innovative fuel cycles \cite{Col18}. In traditional experiments, the direct measurement of neutron-induced cross sections of short-lived nuclei is very challenging because of the difficulties to produce and handle radioactive targets. Performing the same reaction in inverse kinematics, with the heavy, radioactive nucleus impinging upon a target of neutrons, is not possible either, since free neutron targets are currently not available. For these reasons, when the target nuclei are highly radioactive, experimental data are scarce and most of the neutron-induced reaction cross sections rely on theoretical model predictions. However, these predictions often have large uncertainties due to difficulties in describing the de-excitation process of the nucleus formed after the capture of the neutron. This process is ruled by fundamental properties ($\gamma$-ray strength functions, nuclear level densities, fission barriers, etc.) for which the existing nuclear models give very different predictions. This can lead to discrepancies between the calculated cross sections as large as two orders of magnitude or more \cite{Arn07, Lid16}.\\
Indirect methods have been developed to infer neutron-induced reaction cross sections of short-lived nuclei \cite{Bau96, Typ03, Esc12, Lar19, Esc16}. Here we use the surrogate reaction method \cite{Esc12}. In this method, the excited nucleus produced in the neutron-induced reaction of interest is formed through an alternative and experimentally feasible binary reaction, typically an inelastic scattering or a transfer reaction. The measurement of the probabilities of the different decay channels of the excited nucleus ($\gamma$-ray emission, neutron emission, fission, etc.) as a function of its excitation energy provides the information which is required to constrain the models of the above-mentioned nuclear properties. This significantly improves the predictions of the cross sections of the neutron-induced reactions of interest. To date, the surrogate-reaction method has been used and successfully benchmarked in direct kinematics, see e.g. \cite{Esc12, Esc18, Rat18, San20}. \\
The probability $P_{\chi}$ that a nucleus with excitation energy $E^{*}$ formed with a surrogate reaction $X(a,b)$ decays via channel $\chi$ is given by the expression:\\
\begin{equation}
 P_{\chi}(E^{*})=\frac{N_{c,\chi}(E^{*})}{N_{s}(E^{*})\cdot\epsilon_{\chi}(E^{*})}~,%
\end{equation}
where $N_{s}$ is the number of light ejectiles $b$ measured, the so-called single events. $N_{c,\chi}$ is the number of products of decay channel $\chi$ measured in coincidence with the ejectiles $b$ and $\epsilon_{\chi}$ is the efficiency for detecting the products of decay $\chi$ for the reactions in which the outgoing ejectile $b$ is detected. The excitation energy $E^{*}$ is obtained by measuring the kinetic energies of the projectile beam and of the ejectile $b$, and the angle $\theta$ between them.\\
Surrogate-reaction experiments in direct kinematics (where the light nucleus $a$ is the projectile and the heavy nucleus $X$ is at rest) have significant limitations. When the nuclei of interest are far from stability, the targets required for the surrogate reaction are also unavailable. Additionally, competing reactions in target contaminants (such as oxygen) and backings produce a large background, which is very complicated or even impossible to remove \cite{Kes15}. Furthermore, the heavy products of the decay of the excited nucleus are stopped in the target and cannot be detected. Therefore, the measurement of $\gamma$- and neutron-emission probabilities requires detecting the emitted $\gamma$ rays and neutrons. However, the $\gamma$-ray-cascade detection efficiencies in surrogate-reaction experiments are limited to about 20 $\%$ \cite{San19}. The measurement of the neutron-emission probability is extremely challenging and to our knowledge has never been accomplished. \\
Some of the latter limitations can be solved by using the surrogate-reaction method in inverse kinematics, which enables the formation of short-lived nuclei by using a radioactive ion beam and the detection of the heavy, beam-like residues produced after the emission of $\gamma$-rays and neutrons. However, the decay probabilities change very rapidly with excitation energy at the neutron-emission and at the fission thresholds, see e.g. \cite{San20}. The excitation-energy resolution required to scan this rapid evolution is a few 100 keV (FWHM), which is quite difficult to achieve for heavy nuclei in inverse kinematics, due to so far unresolved target issues. Indeed, the required large target density and thickness result in significant energy loss and straggling effects that translate into a large uncertainty in the energies of the projectile and the target-like residue, and in the emission angle $\theta$ at the interaction point. In addition, the presence of target windows and impurities induces background.\\
Here we address these target issues by investigating for the first time surrogate reactions at a heavy-ion storage ring \cite{Ste20}. A key component of storage rings is the electron cooler, which significantly reduces the size, angular divergence and energy spread of the revolving ion beam. If a gas target is present in the ring, the electron cooler compensates for the energy loss and for the energy and angular straggling of the beam taking place during each passage of the beam through the target. As a result, the ion beam always reaches the target with the same energy and the same outstanding quality, making energy loss and straggling effects in the target negligible. Additionally, the frequent passing of the target zone (about 1 million times per second at a few tens of MeV/nucleon) makes possible the use of pure gas targets with ultra-low density ($\approx10^{13}$ atoms/cm$^{2}$) and no windows are necessary. The very low gas target density makes the probability of two consecutive reactions occurring in the target, a nuclear reaction followed by an atomic reaction and vice versa, extremely low ($\approx10^{-20}$). The beam-like residues resulting from the nuclear reaction will therefore possess the same charge state as the beam. \\
Heavy-ion storage rings have to be operated in ultra-high vacuum (UHV) conditions ($10^{-10}$ to $10^{-11}$ mbar), which poses severe constraints on in-ring detection systems. UHV-compatible silicon detectors have only started to be used for a few years in pioneering in-ring nuclear reaction experiments \cite{Zam17, Glo19, Sch23} at the Experimental Storage Ring (ESR) \cite{Fra87} and the CRYRING storage ring \cite{Bru23} of the GSI/FAIR facility.\\
We have conducted the first surrogate-reaction experiment at the ESR with the aim to use the $^{208}$Pb(p,p’) surrogate reaction to asses theoretical models and provide predictions for the neutron-induced radiative capture cross section (n,$\gamma$) of $^{207}$Pb at neutron energies above 800 keV, where no experimental data are available. These data are important for the design of lead-cooled fast reactors \cite{Dom06}. In our experiment, $^{208}$Pb$^{82+}$ projectiles at 30.77 MeV/nucleon were excited by inelastic scattering reactions with a gas-jet target of hydrogen. We had on average $5\cdot10^{7}$ cooled and decelerated, bare $^{208}$Pb$^{82+}$ ions per measurement cycle, revolving at a frequency of 0.695 MHz. The average target thickness was $6\cdot10^{13}$ atoms/cm$^{2}$. We measured the inelastically scattered protons with a Si $\Delta$E-E telescope and the beam-like residues produced after the de-excitation of $^{208}$Pb* via $\gamma$-ray and neutron emission with a position-sensitive Si-strip detector placed behind the dipole magnet downstream from the target (denoted beam-like residue detector in Fig. 1). The dipole separated the unreacted beam, the $^{208}$Pb$^{82+}$ residues produced after $\gamma$-ray emission and the $^{207}$Pb$^{82+}$ residues produced after neutron emission, see Fig. 1. \\
\begin{figure}
\includegraphics[scale=0.45]{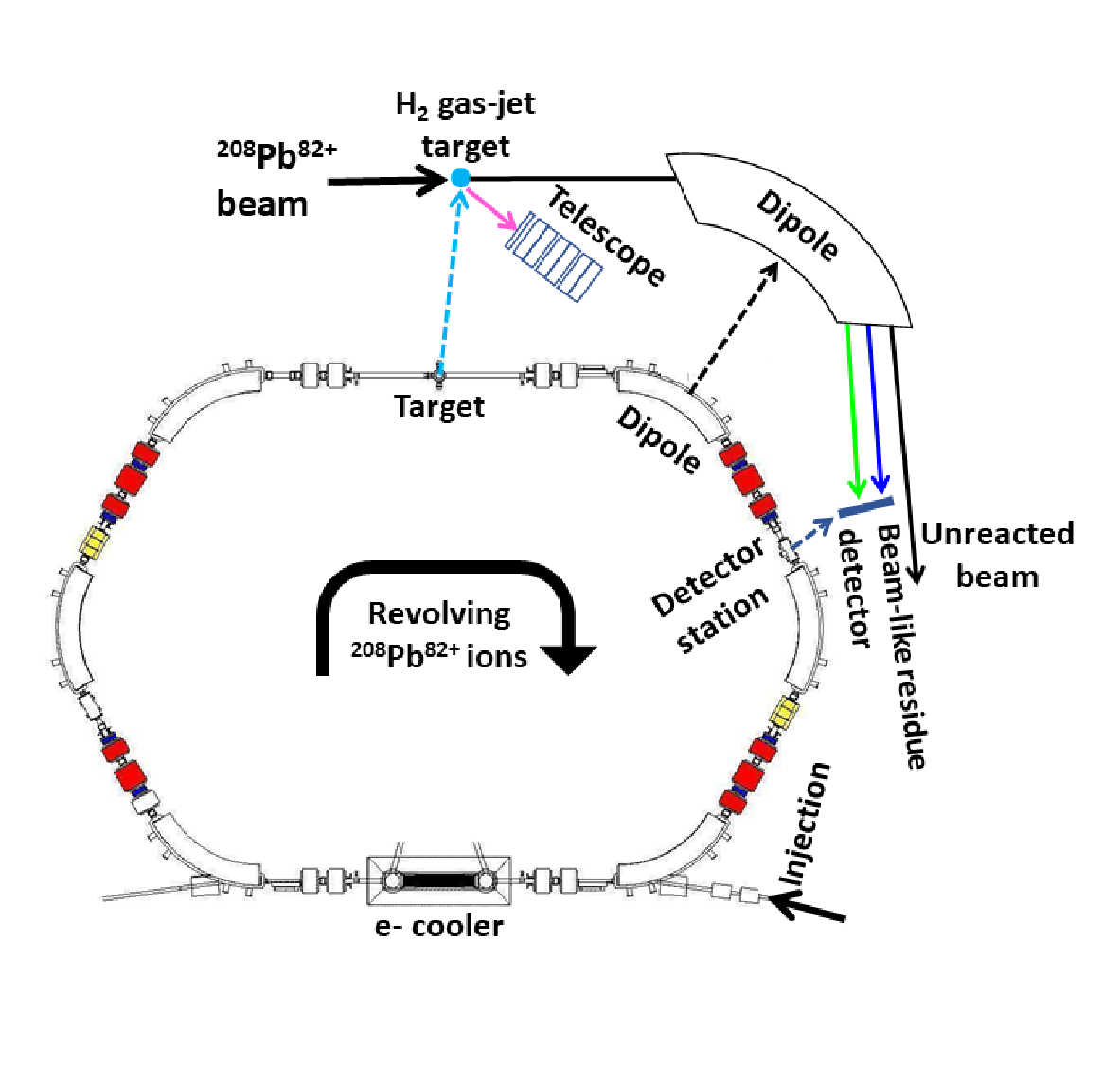}
\caption{ The lower part shows a schematic view of the ESR. The upper part shows the portion of the ring where our detectors have been installed. The trajectories of the scattered protons, the beam, the $^{208}$Pb$^{82+}$ residues produced after $\gamma$ emission and the $^{207}$Pb$^{82+}$ residues formed after neutron emission are represented by the solid pink, black, blue and green arrows, respectively.}
\end{figure}
To prevent detector components from degrading the UHV of the ring, the telescope and the beam-like residue detector were housed in pockets behind 25 $\mu$m thin stainless-steel windows  through which the scattered protons and the heavy beam residues could pass. The telescope was placed at $60^{\circ}$ with respect to the beam axis, at a distance of 10.13 cm from the target. The $\Delta$E detector of the telescope consisted of a 530 $\mu$m-thick double-sided silicon-strip detector (DSSD) of 20×20 mm$^2$ with 16 vertical and 16 horizontal strips, which enabled the measurement of the angle $\theta$ within the angular range from 54.8 to 64.6$^{\circ}$. The angular resolution was estimated to be 0.2$^{\circ}$ (RMS), assuming isotropic emission of the target residues from the center of the target. The $\Delta$E detector was followed by a stack of six single area Si detectors for full energy measurements. Each of the latter E detectors had an active area of 20×20 mm$^2$ and a thickness of 1.51 mm. In inverse kinematics it is possible to have two kinematic solutions leading to two groups of ejectiles having different kinetic energies, but the same angle $\theta$ \cite{Prepa}. In our experiment, scattered protons from the first kinematic solution with kinetic energies above 9.2 MeV passed through the $\Delta$E detector, while scattered protons from the second kinematic solution with kinetic energies between 2.5 and 9.2 MeV were stopped in the $\Delta$E detector, see \cite{Prepa}. The beam-like residue detector was a DSSD with a thickness of 500 $\mu$m, an active area of 122×40 mm$^{2}$, 122 vertical strips and 40 horizontal strips. It was positioned 15.0$\pm$0.1 mm from the beam axis. With this distance we ensured that the rate of elastic scattered beam ions over the whole detector was well within the radiation-damage tolerance range of the detector, which remained operational throughout the experiment. \\
Figure 2 shows the position spectrum of beam-like residues detected in coincidence with scattered protons detected in the telescope. In panel (a) we see the heavy residues measured in coincidence with protons from the first kinematic solution. We can clearly distinguish two peaks; the left peak contains the $^{208}$Pb$^{82+}$ nuclei formed after $\gamma$ emission and the right peak the $^{207}$Pb$^{82+}$ nuclei produced after neutron emission. In panel (b) are shown the heavy residues detected in coincidence with protons from the second kinematic solution for $E^*$=6.5-9.1 MeV and $\theta=56.1-60.40^{\circ}$. In this case, the beam-like residues have larger kinetic energies and their trajectories after the dipole magnet are closer to the beam axis. The $^{208}$Pb$^{82+}$ residues formed after $\gamma$ emission cannot be detected, but all the trajectories of the $^{207}$Pb$^{82+}$ residues formed after neutron emission impinge on the beam-like detector, leading to the observed peak in the position spectrum. We emphasize that in this experiment the efficiency $\epsilon_n$ for the neutron emission channel is 100$\%$ \cite{Prepa}. The largest loss of efficiency comes from electron capture of the $^{207}$Pb$^{82+}$ residues in the residual gas between the target and the beam-like detector. The probability for this event has been calculated to account only to $\approx10^{-10}$, so it can be neglected.\\
\begin{figure}
\includegraphics[scale=0.45]{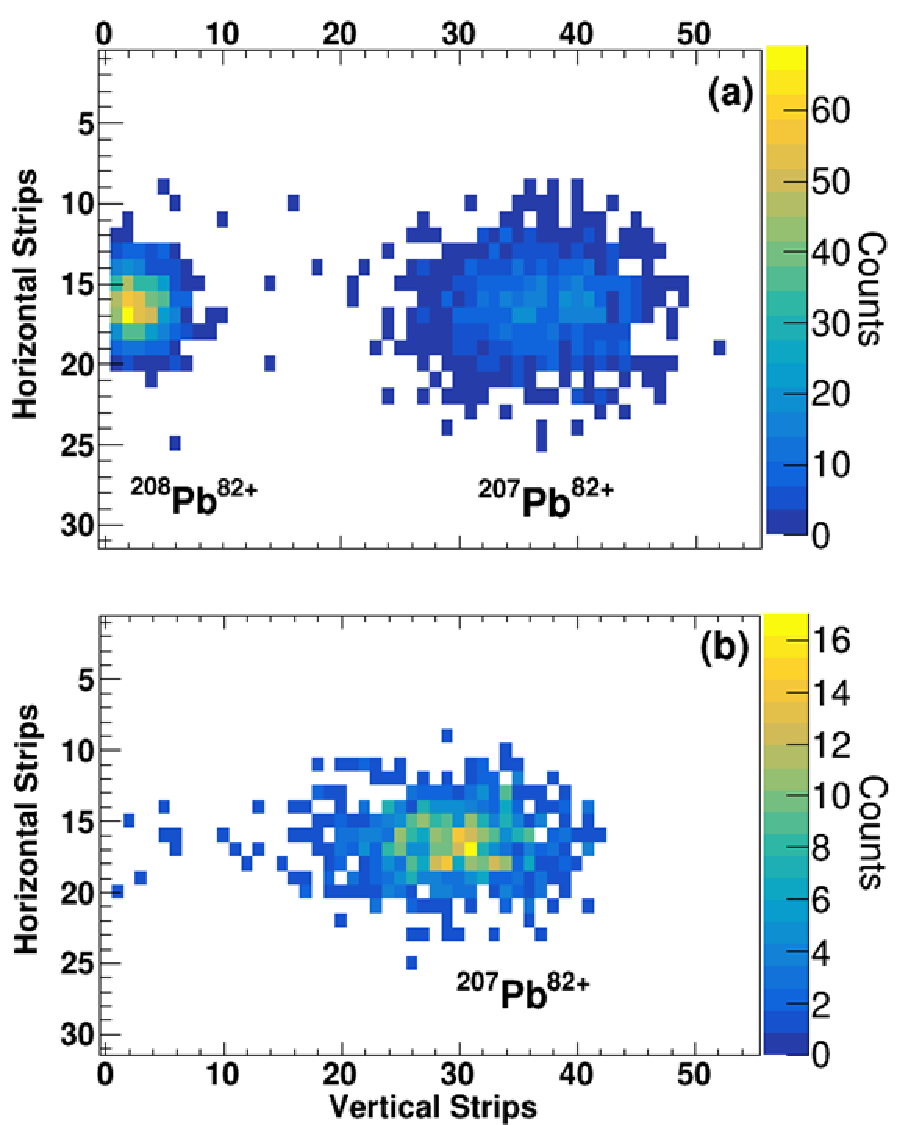}
\caption{Position of beam-like residues measured in coincidence with detected scattered protons from the first (a) and the second kinematic solution (b), see text for details.  }
\end{figure}
In this work, we only consider the results obtained with the second kinematic solution, the results of the first kinematic solution are discussed in \cite{Prepa}. We obtained the singles spectrum $N_s(E^*)$ by representing the number of detected protons as a function of the $E^*$ of $^{208}$Pb. The coincidence spectrum $N_{c,n}(E^*)$ was inferred by representing the number of protons detected in coincidence with the beam-like residues located within the peak of Fig. 2 (b). The bin size of these two histograms was 200 keV. By computing the ratio of $N_{c,n}(E^*)$ over $N_s(E^*)$ and using $\epsilon_n=1$ (see eq. (1)), we were able to measure for the first time the neutron-emission probability $P_{n}(E^*)$, as illustrated in Fig. 3. The displayed error bars include the covariance between $N_{c,n}(E^*)$ and $N_s(E^*)$ \cite{Prepa}. Thanks to the 100$\%$ detection efficiency for the heavy residues, it has been possible to achieve relative uncertainties of less than 6$\%$, despite the small total number of 1581 single events measured. The experimental data show an onset of $P_n$ below the neutron separation energy $S_n$. As discussed below, this is due to the excitation energy resolution $\Delta E^*$, which is $\approx$ 240 keV (RMS). We estimated $\Delta E^*$ with a simulation, which was benchmarked with the well-separated ground-state peak of $^{208}$Pb at $E^*$=0 MeV, see \cite{Prepa}. In this experiment, $\Delta E^*$ is dominated by the uncertainty in the proton scattering angle $\theta$ induced by the target radius of 2.5 mm.\\
\begin{figure}
\includegraphics[scale=0.48]{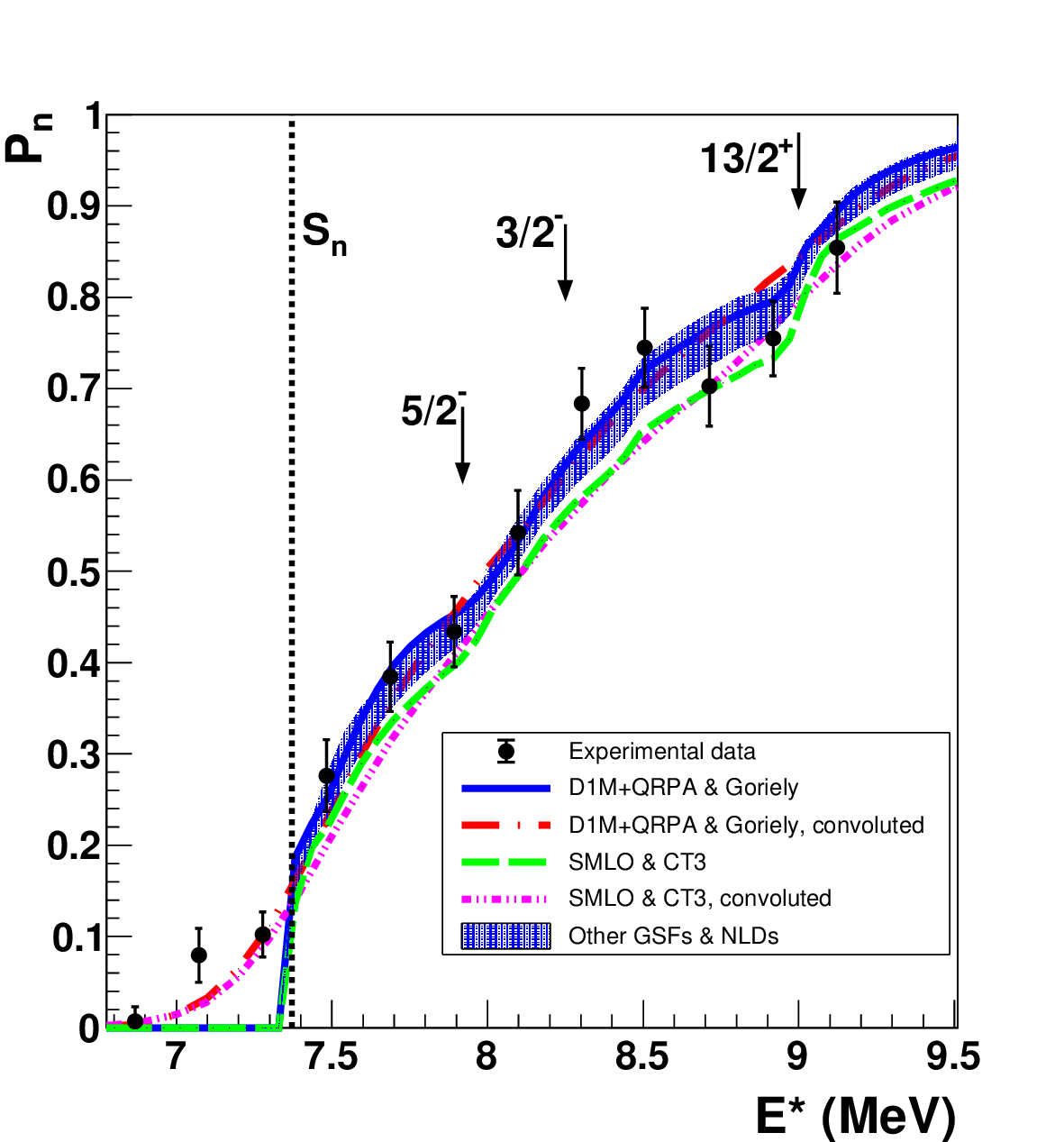}
\caption{ Neutron-emission probability as a function of the excitation energy $E^*$ of $^{208}$Pb measured for the $^{208}$Pb(p,p’) reaction in comparison with TALYS calculations. The arrows indicate the $E^*$ at which the three first excited states of $^{207}$Pb become accessible. The spin and parity of the states are also given. The neutron separation energy $S_n$ of $^{208}$Pb at $E^*$=7.37 MeV is indicated by the vertical dotted line, see text for details. }
\end{figure}
To compare our results with theory, we have calculated $P_n(E^*)$ with the statistical model using the expression:
\begin{equation}
 P_{n}(E^{*})=\sum_{J^{\pi}}F(E^{*},J^{\pi})\cdot G_{n}(E^{*},J^{\pi})~,%
\end{equation}
where $F(E^{*},J^{\pi})$ is the probability to form the excited nucleus in a state of spin $J$ and parity $\pi$ at an excitation energy $E^*$ by the $^{208}$Pb(p,p’) reaction, and $G_{n}(E^{*},J^{\pi})$ is the probability that the nucleus decays from that state via neutron emission. The $J^\pi$ distributions given by $F$ were calculated with the microscopic description developed in \cite{Dup06, Dup19}. The theoretical formalism and the results for $F(J^\pi)$ at $E^*= 8$ and 9 MeV are presented in the Supplemental Material \cite{supmat}. To determine $G_n$ we used the statistical Hauser-Feshbach model of TALYS 1.96 \cite{talys}. Among all the quantities needed to describe the de-excitation of $^{208}$Pb, the $\gamma$-ray strength function (GSF) and the nuclear level density (NLD) are the most uncertain ones. We considered different models for these two quantities with adjusted parameters for $^{208}$Pb, which we obtained from literature. For the GSF, we utilized three models: the model of Kopecky and Uhl \cite{Kop90}, the Simple Modified Lorentzian model (SMLO) \cite{Gor19} and the results of Hartree-Fock-Bogolyubov (HFB) and Quasi-particle Random Phase Approximation (QRPA) calculations based on the Gogny D1M nuclear interaction \cite{Gor18}, which we will denote as D1M+QRPA. Regarding the NLD, we employed six distinct descriptions, three of these were based on the constant-temperature (CT) model \cite{Gil65} with different adjusted parameters, they are denoted CT1, CT2 and CT3. One description was based on the back-shifted Fermi-gas (BSFG) model \cite{Dil73}. The two others were the microscopic NLDs by Goriely et al. \cite{Gor08} and Hilaire et al. \cite{Hil12}. Further details on the models and used parameters can be found in \cite{supmat}.\\
%
%
We combined the three GSF descriptions with the six NLD models leading to 18 different TALYS calculations. We expect to observe significant differences between the calculations and our data at $S_{n}$ due to the excitation energy resolution $\Delta E^*$. To account for $\Delta E^*$ we convoluted the calculations with a Gaussian function with a standard deviation of 240 keV. We have evaluated the deviations between the calculations and our data by computing the reduced ${\chi}^2$ before and after the convolution, see Table II in \cite{supmat}. The deviations decrease drastically after the convolution for all the calculations. The calculations that use the NLD CT3 and the calculation utilising the SMLO and the BSFG models have a reduced ${\chi}^2$ exceeding 1.63 and can be excluded by our data with a confidence level of 93$\%$. These calculations are also excluded by the data from the first kinematic solution, see \cite{Prepa}. We obtained the best agreement (lowest residuals and reduced ${\chi}^2$, see \cite{supmat}) with the convoluted calculation using the D1M+QRPA GSF model and the NLD by Goriely et al. The latter calculation is compared with our data before and after convolution in Fig. 3. As shown by the red dashed-dotted line, the convoluted result exhibits a significantly improved agreement with the experimental data below $S_n$. The calculation obtained with the SMLO and the CT3 models yields the largest reduced ${\chi}^2$. Between $S_n$ and 8.5 MeV, this calculation is systematically below our data, the best TALYS calculation and the blue shaded area, which includes all the calculations except those that are excluded by our data, see Fig. 3. The calculations show an increase at $E^*\approx 8$ and 9 MeV. These increases occur when the $E^*$ of the $^{207}$Pb residue formed after neutron emission is high enough to populate the 1$^{st}$ and 3$^{rd}$ excited states of $^{207}$Pb, see the arrows in Fig. 3. The population of the 3$^{rd}$ excited state is particularly favoured because its spin $J$=13/2 is closer to the spins populated in the $^{208}$Pb(p,p') reaction (average spin $\overline{J}\approx 5.4$, see \cite{supmat}).
\begin{figure}[h]
\includegraphics[scale=0.48]{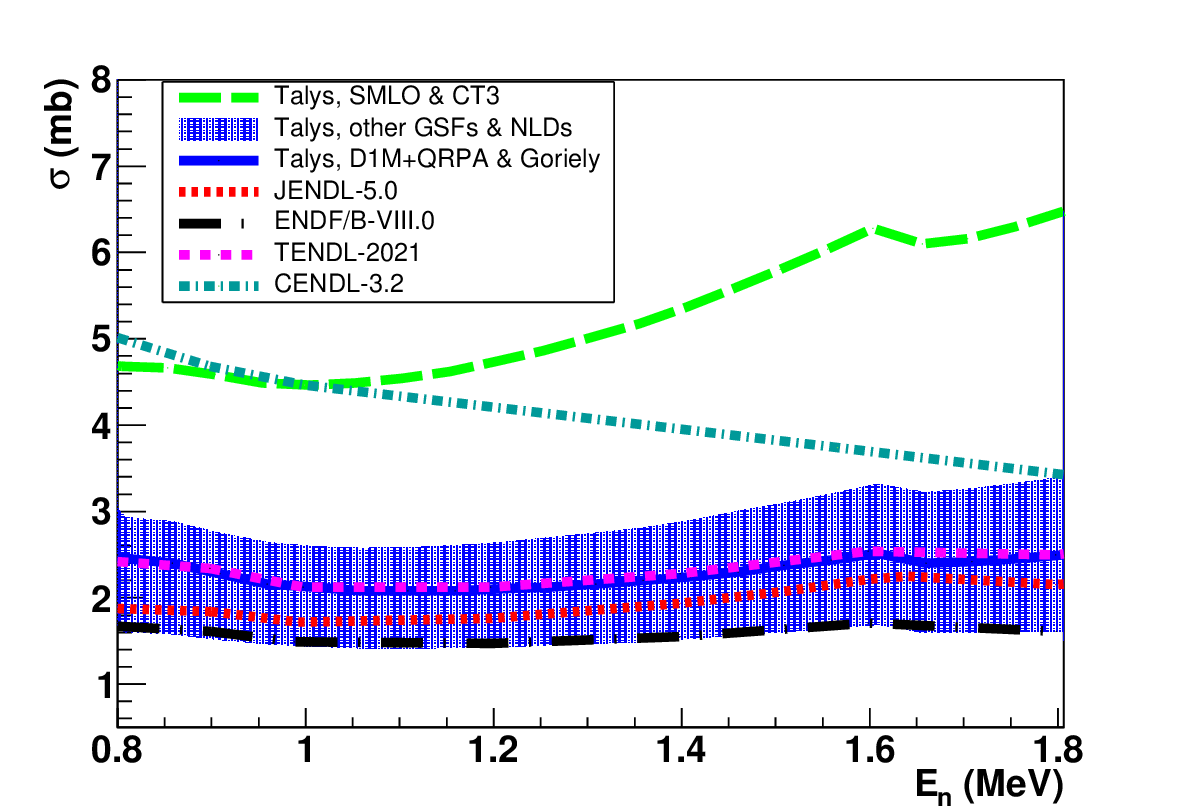}
\caption{ $^{207}Pb(n,\gamma)$ cross section as a function of neutron energy $E_n$. The results of our TALYS calculations are compared with the JENDL-5.0 \cite{Iwa23}, TENDL-2021 \cite{Kon19}, CENDL-3.2 \cite{Ge20} and ENDF/B-VIII.0 \cite{Bro18} evaluations, see text for details.}
\end{figure}
\\
We have used the GSF and NLD models that are not excluded by our data to calculate the $^{207}$Pb(n,$\gamma$) cross section above neutron energies of 800 keV, see blue line and shaded area in Fig. 4. Above 800 keV, several evaluations (i.e. databases of recommended values for the cross sections \cite{Ber19}) based on the Hauser-Feshbach formalism are available to which we can compare our results. We expect that the calculation obtained with the SMLO and the CT3 models will result in a larger cross section, as this calculation leads to lower values of $P_n(E^*)$ and thus higher values of the $\gamma$-emission probability, since in the covered $E^*$ range $\gamma$ and neutron emission are the only open decay channels. The green dashed line in Fig. 4 shows that this is indeed the case, this TALYS calculation is well above all the other TALYS calculations. This demonstrates the strong connection between $P_n(E^*)$ and the (n, $\gamma$) cross section, and the usefulness of employing the $P_n(E^*)$ from a surrogate reaction for constraining predictions for radiative capture cross sections. As shown in Fig. 4, our calculations encompass all the evaluations except CENDL-3.2 \cite{Ge20}, which shows a very different shape and is above our results. \\
In conclusion, we have measured for the first time the neutron emission probability as a function of the excitation energy, $P_n(E^*)$, of $^{208}$Pb. Our measurement benefited from the unrivaled advantages of the ESR heavy-ion storage ring, which allowed us to detect the beam-like residues formed after neutron emission with an efficiency of 100$\%$. We employed our results for $P_n(E^*)$ to select various combinations of models for the $\gamma$-ray strength function and the nuclear level density of $^{208}$Pb available in the literature. The selected models were used to infer the $^{207}$Pb(n,$\gamma$) cross section at neutron energies for which no experimental data are available. This demonstrates the advantage of using the $P_n(E^*)$ obtained through surrogate-reaction experiments to constrain predictions for (n,$\gamma$) cross sections. Our results are in good agreement with the JENDL-5.0, TENDL-2021 and ENDF/B-VIII.0 evaluations, but disagree with the CENDL-3.2 evaluation. In the future, we will complete our setup with fission detectors to measure also the fission probabilities, increase the solid angle of the telescope and use a target with a smaller radius, which will allow us to improve the excitation energy resolution. With these improvements we will be able to conduct next-generation experiments with radioactive stored beams, where we will measure simultaneously and with high precision the probabilities for all the de-excitation channels (fission, $\gamma$, neutron and two-neutron emission) of many short-lived nuclei for which the neutron-induced cross sections are considered impossible to measure.\\
\\
This work is supported by the European Research Council (ERC) under the European Union’s Horizon 2020 research and innovation programme (ERC-Advanced grant NECTAR, grant agreement No 884715). The results presented here are based on the experiment E146, which was performed at the ESR storage ring of the GSI Helmholtzzentrum fuer Schwerionenforschung, Darmstadt (Germany) in the context of FAIR Phase-0. We thank the Prime 80 program from the CNRS for funding the PhD thesis of MS and the GSI/IN2P3 collaboration 19-80. JG, YuAL, RR and ThS acknowledge support by the State of Hesse within the Cluster Project ELEMENTS (Project ID 500/10.006). AH is grateful for funding from the Knut and Alice Wallenberg Foundation under KAW 2020.0076.


\bibliography{maintext}

\end{document}


\preprint{AIP/123-QED}

\title{Supplemental Material for: First measurement of the neutron-emission probability with a surrogate reaction in inverse kinematics at a heavy-ion storage ring\\}

\author{M. Sguazzin}
\thanks{Present address: IJCLab, 91405 Orsay, France}
\affiliation{Université de Bordeaux, CNRS, LP2I Bordeaux, 33170 Gradignan, France}
\author{B. Jurado}
\thanks{Corresponding author, jurado@cenbg.in2p3.fr}
\affiliation{Université de Bordeaux, CNRS, LP2I Bordeaux, 33170 Gradignan, France}
\author{J. Pibernat}
\affiliation{Université de Bordeaux, CNRS, LP2I Bordeaux, 33170 Gradignan, France}
\author{J. A. Swartz}
\thanks{Present address: FRIB, MSU, Michigan 48824, USA}
\affiliation{Université de Bordeaux, CNRS, LP2I Bordeaux, 33170 Gradignan, France}
\author{M. Grieser}
\affiliation{Max-Planck-Institut für Kernphysik, 69117 Heidelberg, Germany}
\author{J. Glorius}
\affiliation{GSI Helmholtzzentrum für Schwerionenforschung, 64291 Darmstadt, Germany}
\author{\mbox{Yu. A. Litvinov}}
\affiliation{GSI Helmholtzzentrum für Schwerionenforschung, 64291 Darmstadt, Germany}
\author{\mbox{J. Adamczewski-Musch}}
\affiliation{GSI Helmholtzzentrum für Schwerionenforschung, 64291 Darmstadt, Germany}
\author{P. Alfaurt}
\affiliation{Université de Bordeaux, CNRS, LP2I Bordeaux, 33170 Gradignan, France}
\author{P. Ascher}
\affiliation{Université de Bordeaux, CNRS, LP2I Bordeaux, 33170 Gradignan, France}
\author{L. Audouin}
\affiliation{Université Paris-Saclay, CNRS, IJCLab, 91405 Orsay, France}
\author{C. Berthelot} 
\affiliation{Université de Bordeaux, CNRS, LP2I Bordeaux, 33170 Gradignan, France}
\author{B. Blank}
\affiliation{Université de Bordeaux, CNRS, LP2I Bordeaux, 33170 Gradignan, France}
\author{\mbox{K. Blaum}}
\affiliation{Max-Planck-Institut für Kernphysik, 69117 Heidelberg, Germany}
\author{\mbox{B. Br\"uckner}}
\affiliation{Goethe University of Frankfurt, 60438 Frankfurt, Germany}
\author{\mbox{S. Dellmann}}
\affiliation{Goethe University of Frankfurt, 60438 Frankfurt, Germany}
\author{I. Dillmann}
\affiliation{TRIUMF, Vancouver, British Columbia, V6T 2A3, Canada}
\affiliation{Department of Physics and Astronomy, University of Victoria, Victoria, BC, V8P 5C2, Canada}
\author{C. Domingo-Pardo} 
\affiliation{IFIC, CSIC-Universidad de Valencia, 46980 Valencia, Spain}
\author{\mbox{M. Dupuis}} 
\affiliation{CEA, DAM, DIF, 91297 Arpajon, France}
\affiliation{Université Paris-Saclay, CEA, LMCE, 91680 Bruyères-Le-Châtel, France}
\author{\mbox{P. Erbacher}}
\affiliation{Goethe University of Frankfurt, 60438 Frankfurt, Germany}
\author{M. Flayol} 
\affiliation{Université de Bordeaux, CNRS, LP2I Bordeaux, 33170 Gradignan, France}
\author{\mbox{O. Forstner}}
\affiliation{GSI Helmholtzzentrum für Schwerionenforschung, 64291 Darmstadt, Germany}
\author{\mbox{D. Freire-Fernández}}
\affiliation{Max-Planck-Institut für Kernphysik, 69117 Heidelberg, Germany}
\affiliation{Ruprecht-Karls-Universität Heidelberg, 69117 Heidelberg, Germany}
\author{M. Gerbaux}
\affiliation{Université de Bordeaux, CNRS, LP2I Bordeaux, 33170 Gradignan, France}
\author{\mbox{J. Giovinazzo}} 
\affiliation{Université de Bordeaux, CNRS, LP2I Bordeaux, 33170 Gradignan, France}
\author{\mbox{S. Grévy}}
\affiliation{Université de Bordeaux, CNRS, LP2I Bordeaux, 33170 Gradignan, France}
\author{C. J. Griffin}
\affiliation{TRIUMF, Vancouver, British Columbia, V6T 2A3, Canada}
\author{A. Gumberidze}
\affiliation{GSI Helmholtzzentrum für Schwerionenforschung, 64291 Darmstadt, Germany}
\author{\mbox{S. Heil}} 
\affiliation{Goethe University of Frankfurt, 60438 Frankfurt, Germany}
\author{\mbox{A. Heinz}}
\affiliation{Chalmers University of Technology, 41296 Gothenburg, Sweden}
\author{R. Hess}
\affiliation{GSI Helmholtzzentrum für Schwerionenforschung, 64291 Darmstadt, Germany}
\author{D. Kurtulgil} 
\affiliation{Goethe University of Frankfurt, 60438 Frankfurt, Germany}
\author{\mbox{N. Kurz}}
\affiliation{GSI Helmholtzzentrum für Schwerionenforschung, 64291 Darmstadt, Germany}
\author{G. Leckenby}
\affiliation{TRIUMF, Vancouver, British Columbia, V6T 2A3, Canada}
\affiliation{Department of Physics and Astronomy, University of British Columbia, Vancouver, BC, V6T 1Z1, Canada}
\author{S. Litvinov} 
\affiliation{GSI Helmholtzzentrum für Schwerionenforschung, 64291 Darmstadt, Germany}
\author{B. Lorentz}
\affiliation{GSI Helmholtzzentrum für Schwerionenforschung, 64291 Darmstadt, Germany}
\author{V. Méot} 
\affiliation{CEA, DAM, DIF, 91297 Arpajon, France}
\affiliation{Université Paris-Saclay, CEA, LMCE, 91680 Bruyères-Le-Châtel, France}
\author{\mbox{J. Michaud}}
\thanks{Present address: IJCLab, 91405 Orsay, France}
\affiliation{Université de Bordeaux, CNRS, LP2I Bordeaux, 33170 Gradignan, France}
\author{\mbox{S. Pérard}}
\affiliation{Université de Bordeaux, CNRS, LP2I Bordeaux, 33170 Gradignan, France}
\author{\mbox{N. Petridis}}
\affiliation{GSI Helmholtzzentrum für Schwerionenforschung, 64291 Darmstadt, Germany}
\author{U. Popp}
\affiliation{GSI Helmholtzzentrum für Schwerionenforschung, 64291 Darmstadt, Germany}
\author{D. Ramos}
\affiliation{GANIL, CRNS/IN2P3-CEA/DRF, 14000 Caen, France}
\author{R. Reifarth} 
\affiliation{Goethe University of Frankfurt, 60438 Frankfurt, Germany}
\affiliation{Los Alamos National Laboratory, Los Alamos, NM, 87544, USA}
\author{M. Roche}
\affiliation{Université de Bordeaux, CNRS, LP2I Bordeaux, 33170 Gradignan, France}
\author{M.S. Sanjari} 
\affiliation{GSI Helmholtzzentrum für Schwerionenforschung, 64291 Darmstadt, Germany}
\affiliation{Aachen University of Applied Sciences, Aachen, Germany}
\author{\mbox{R.S. Sidhu}} 
\affiliation{School of Physics and Astronomy, University of Edinburgh, EH9 3FD Edinburgh, United Kingdom}
\affiliation{GSI Helmholtzzentrum für Schwerionenforschung, 64291 Darmstadt, Germany}
\affiliation{Max-Planck-Institut für Kernphysik, 69117 Heidelberg, Germany}
\author{\mbox{U. Spillmann}} 
\affiliation{GSI Helmholtzzentrum für Schwerionenforschung, 64291 Darmstadt, Germany}
\author{M. Steck}  
\affiliation{GSI Helmholtzzentrum für Schwerionenforschung, 64291 Darmstadt, Germany}
\author{Th. Stöhlker} 
\affiliation{GSI Helmholtzzentrum für Schwerionenforschung, 64291 Darmstadt, Germany}
\author{B. Thomas}
\affiliation{Université de Bordeaux, CNRS, LP2I Bordeaux, 33170 Gradignan, France}
\author{L. Thulliez}
\affiliation{IRFU, CEA, Université Paris-Saclay, 91191 Gif-sur-Yvette, France}
\author{M. Versteegen}
\affiliation{Université de Bordeaux, CNRS, LP2I Bordeaux, 33170 Gradignan, France}
\author{B. Włoch}
\affiliation{Université de Bordeaux, CNRS, LP2I Bordeaux, 33170 Gradignan, France}


\date{\today}
\maketitle

\section{Angular momentum and parity distributions populated in the $^{208}Pb(p,p')$ reaction}

The angular momentum and parity distributions $F(E^*, J^\pi)$ were calculated with the microscopic description of \cite{Dup06, Dup19}. In this approach, the excited states of $^{208}$Pb in the continuum are determined with the Random Phase Approximation (RPA) using the Gogny D1S interaction \cite{Ber89}. The distorted-wave Born approximation is used to determine the cross sections to populate these excited states with the $^{208}$Pb(p,p’) reaction. The Jeukenne, Lejeune, and Mahaux (JLM) microscopic optical model potential \cite{Dup19} is employed for the population of natural parity states and the Melbourne microscopic optical model \cite{Dup06} is used for non-natural parity states. \\

The JLM model incorporates rearrangement corrections and effectively reproduces the cross sections for collective state excitations at incident energies of 30 MeV. The less collective nature of non-natural parity excitations reduces the significance of rearrangement corrections. However, the description of non-natural parity transitions requires considering two-body spin-orbit and tensor interactions, which are included in the Melbourne g-matrix model but not in the JLM model. The combination of these two models thus ensures a comprehensive coverage of the different possible excitations in proton inelastic scattering reactions. Further details will be provided in an upcoming publication \cite{Dupfuture}.\\

In this work, we use the Melbourne and the JLM folding models with nuclear structure input from RPA for predicting cross sections up to relatively high excitation energies $E^*$ of 9 MeV. The accuracy of differential cross sections primarily depends on the nuclear structure inputs used. The Melbourne folding model has already been used to describe inelastic scattering to discrete levels in $^{208}$Pb for non-natural parity excitations ($E^*$=5-7.5 MeV) and giant resonances ($E^*$=10-21 MeV) \cite{Dupthesis}, as well as proton scattering on $^{90}$Zr for excitations in the energy range $E^*$=0-20 MeV \cite{Dup17}. The JLM folding model has been extensively used for neutron and proton inelastic scattering across a wide range of incident energies (from threshold up to 200 MeV) and various targets, see e.g. \cite{Dup19}. Although there are no published applications using the JLM folding model for high-energy RPA excitations that compare inelastic cross sections to experimental data, the studies conducted with the Melbourne model and alternative folding models cover a wide range of excitation energies and validate the use of RPA structures for high-energy excitations in inelastic scattering. In addition, the JLM folding model with RPA and QRPA (Quasi-particle Random Phase Approximation) excitations has been applied for neutron inelastic scattering off various targets with mass A=16 to 240 for incident neutron energies below 30 MeV to predict the spin distributions of residual nuclei, which may decay by the emission of a $\gamma$-ray cascade. The predicted spin distributions lead to a good reproduction of (n,n$^{,}\gamma$) cross sections of  $^{238}$U \cite{Ker21}, $^{232}$Th \cite{Partythesis} and $^{182,184,186}$W \cite{Henningfuture}. These results provide an additional element of confidence in the reliability of the JLM model at higher $E^*$.\\

We have calculated the $F(E^*, J^\pi)$ distributions from $E^*= 1.0$ to 9.5 MeV in steps of 0.5 MeV. The distributions $F(E^*, J^\pi)$ vary smoothly with $E^*$ and this variation is considered in the calculation of the neutron emission probability. In the angular range covered by our data, $\theta=56.1$ to 60.4$^{\circ}$, the calculated $F$ distributions show a very weak dependence with the proton scattering angle in the center-of-mass $\theta_{cm}$. The results for $F(J^\pi)$ at $E^*=$ 8 and 9 MeV are shown in Fig. S1. At $E^*=8$ MeV the average spin is 5.8 $\hbar$ for positive and 5.3 $\hbar$ for negative parities, while at $E^*=9$ MeV the average spins are 5.3 $\hbar$ and 5.5 $\hbar$ for positive and negative parities, respectively.\\

\renewcommand{\thefigure}{S1}
\begin{figure}[h]
\includegraphics[scale=0.54]{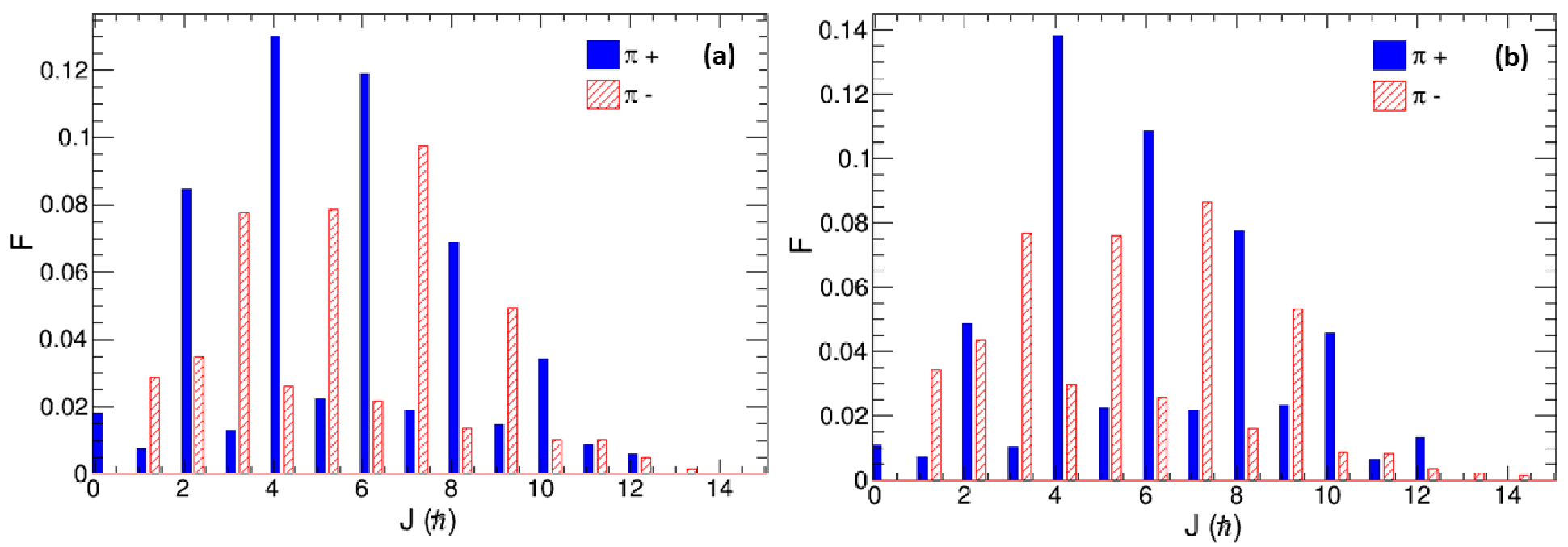}
\caption{\label{fig:epsart} Calculated spin-parity ($J^\pi$) distributions $F$ of $^{208}$Pb populated by the $^{208}$Pb(p, p’) reaction for the center-of-mass angular range $\theta_{cm}=149-$167$^{\circ}$, which corresponds to the angular range in the laboratory $\theta=56.1-$ 60.4$^{\circ}$. The results obtained for excitation energy $E^*$ = 8 MeV and $E^*$ = 9 MeV are shown in panels (a) and (b), respectively. }
\end{figure}

\section{Ingredients for the determination of the de-excitation probabilities $G_{n}$(E*,$J^{\pi})$}

Among the different ingredients needed to compute the probabilities $G_{n}$(E*,$J^{\pi})$ of eq. (2), the most important ones are the optical model potential, which determines the neutron transmission coefficients, the nuclear level density (NLD) and the $\gamma$-ray strength function (GSF). We considered three different optical model potentials, the default potential used in TALYS \cite{Kon03} and two parametrizations of the potential developed by B. Morillon and P. Romain \cite{Mor06, Mor07}, which is well suited for the mass and energy regions considered in this work. The resulting neutron emission probabilities do not show significant differences. The reason is that the optical model potential is well known thanks to the existence of good quality data on neutron-induced total and elastic scattering cross sections on $^{207}$Pb. However, the NLD and GSF of $^{208}$Pb are more uncertain.\\
In this work, we have used different descriptions for the NLD and the GSF of $^{208}$Pb, which we found in the literature, see Table I. Three descriptions of the NLD are based on the constant temperature (CT) model \cite{Gil65}, which has two parameters $T$ and $E_{0}$. The values of these two parameters for each description and the references from which they were taken are given in Table I. We also considered the experimental NLD of $^{208}$Pb measured by Bassauer et al. \cite{Bas16}, which is very well described by the back-shifted Fermi-gas (BSFG) model \cite{Dil73}. The other two NLD descriptions are based on microscopic calculations by Goriely et al. \cite{Gor08} and by Hilaire et al. \cite{Hil12}, the results of which are given in tabular form. Goriely et al. use the effective Skyrme BSk14 nucleon-nucleon interaction, whereas Hilaire et al. utilise the D1M Gogny interaction. The BSFG and the CT3 NLDs are above all the other NLDs. In particular, CT3 is 14 times larger than the other NLDs at $S_n$, with the differences increasing with $E^*$.\\

\begin{table*}[h] 
\caption{Descriptions for the level-density and $\gamma$ strength function of $^{208}Pb$ used in this work. The term CT denotes the constant temperature model \cite{Gil65}, BSFG the back-shifted Fermi gas model \cite{Dil73}, KU the Kopecky and Uhl model \cite{Kop90}, SMLO the Simple Modified Lorentzian model (SMLO) \cite{Gor19} and D1M+QRPA the Quasi-particle Random Phase Approximation (QRPA) calculations based on the Gogny D1M nuclear interaction \cite{Gor18}.}
\begin{ruledtabular}
\begin{tabular}{lcdr}
\textrm{Nuclear level density}&
\textrm{$\gamma$-ray strength function}\\
\colrule
CT1, $T$=0.92 MeV and $E_{0}$=1.37 MeV from \cite{Kon08} & KU with parameters from \cite{talys}\\
CT2, $T$=0.82 MeV and $E_{0}$=1.81 MeV from \cite{talys} & SMLO with parameters from \cite{talys} \\
CT3, $T$=0.69 MeV and $E_{0}$=1.67 MeV from \cite{Iwa07}& Microscopic D1M+QRPA \cite{Gor18} \\
BSFG with parameters from \cite{Bas16} & \\
Table from microscopic calculation by Goriely et al. \cite{Gor08} &  \\
Table from microscopic calculation by Hilaire et al. \cite{Hil12} &  \\
\end{tabular}
\end{ruledtabular}
\end{table*}
In regard to the GSF, two analytical descriptions were employed, the model by Kopecky and Uhl (KU) \cite{Kop90} and the Simple Modified Lorentzian model (SMLO) by Goriely et al. \cite{Gor19}. In both cases we used the parameters given by \cite{talys} for these models for $^{208}$Pb. The results obtained with the SMLO model agree fairly well with the experimental data of \cite{Bas16} above 6 MeV. We also considered the results of Hartree-Fock-Bogolyubov (HFB) and Quasi-particle Random Phase Approximation (QRPA) calculations based on the Gogny D1M nuclear interaction \cite{Gor18}, which have been designated as D1M+QRPA. We carried out calculations with and without the so-called upbend, i.e. an increase in the GSF at decreasing energies approaching zero \cite{Gor18}. However, no significant differences were observed in the results for $P_{n}$($E$*).\\

\section{Comparison of experimental data and model calculations}
We have combined the six NLD descriptions with the three GSF models leading to 18 calculations. Each calculation was convoluted with the excitation energy resolution. We compared the convoluted and non-convoluted calculation with our data for $P_{n}$(E*). We have quantified the deviations of each calculation with respect to our data by computing the residuals, which measure systematic deviations with respect to the experimental data:

\begin{equation}
 R=\frac{1}{N}\sum_{i=1}^{N}(P^{c}_{n,i}-P^{m}_{n,i}),%
\end{equation}

and the reduced $\chi^2$, defined as:

\begin{equation}
 \chi^2=\frac{1}{N}\sum_{i=1}^{N}\frac{(P^{c}_{n,i}-P^{m}_{n,i})^2}{\Delta P_{n,i}^2},%
\end{equation}

where $P^{c}_{n,i}$ and $P^{m}_{n,i}$ are the calculated and the measured neutron-emission probabilities, respectively. $\Delta P_{n,i}$ are the experimental uncertainties and $N$ is the number of degrees of freedom. In the present case, no adjustments have been made to the model parameters and $N$ is equal to the number of data points, namely $N$ = 12. The values of  $R$ and $\chi^2$ obtained for each combination with and without convolution with the excitation energy resolution are listed in Table II. We can see that the $\chi^2$ becomes significantly lower after the convolution. The calculations where the reduced $\chi^2$ is above 1.63 can be rejected by our data with a confidence level of $93\%$. The  SMLO and CT3 combination (see the intersection between column 4 and row 4) is rejected with a confidence level of 99.99$\%$. The best agreement with our data is found for the calculation with the D1M+QRPA model for the GSF and the NLD by Goriely et al. \cite{Gor08}, which shows a residual $R$=0.00 and the lowest reduced $\chi^2$ value after convolution, $\chi^2=1.10$, see the intersection between column 6 and row 3.

\begin{table} [h]
\caption{Values of the residuals $R$ and the reduced $\chi^2$ (rounded to the second decimal point) obtained with the different combinations of the nuclear level densities (NLD) and $\gamma$-ray strength function (GSF). For each combination of GSF and NLD we give first the values for the residuals and below the reduced $\chi^2$. On the left we give the value obtained with the calculation not convoluted with the excitation energy resolution and on the right with the convoluted calculation. The term CT denotes the constant temperature model \cite{Gil65}, BSFG the back-shifted Fermi gas model \cite{Dil73}, KU the Kopecky and Uhl model \cite{Kop90}, SMLO the Simple Modified Lorentzian model (SMLO) \cite{Gor19} and D1M+QRPA the Quasi-particle Random Phase Approximation (QRPA) calculations based on the Gogny D1M nuclear interaction \cite{Gor18}.}
\centering\setlength\tabcolsep{4.2pt}\renewcommand\arraystretch{2.8}
  \noindent\makebox[\textwidth]{%
   \begin{threeparttable}
    \begin{tabular}{|l|c|c|c|c|c|c|c|c|c|c|c|c|}
      \hline 
      \multirow{2}{*}{\diagbox[width=\dimexpr \textwidth/8+4\tabcolsep\relax, height=2.065cm]{GSF}{NLD}} & \multicolumn{2}{c|}{CT1} & \multicolumn{2}{c|}{CT2} & \multicolumn{2}{c|}{CT3} & \multicolumn{2}{c|}{BSFG} & \multicolumn{2}{c|}{Goriely} & \multicolumn{2}{c|}{Hilaire} \\\cline{2-13}
      
& \vspace{0.0mm} \shortstack{No \\ Conv } & Conv & \shortstack{No \\ Conv} & Conv & \shortstack{No \\ Conv} & Conv & \shortstack{No \\ Conv} & Conv & \shortstack{No \\ Conv} & Conv & \shortstack{No \\ Conv} & Conv  \\
\hline                  
KU       & \makecell{-0.01 \\ 2.55}& \makecell{-0.01 \\ 1.14}& \makecell{ -0.01 \\ 2.59}  & \makecell{0.00 \\ 1.14} &  \makecell{-0.04 \\ 3.03}  &  \makecell{-0.03 \\ 1.85}  &  \makecell{-0.02 \\ 2.61} & \makecell{-0.02 \\ 1.31} & \makecell{0.00 \\ 2.63}  & \makecell{0.00 \\ 1.12}  & \makecell{0.00 \\ 2.66} &  \makecell{0.00 \\ 1.18}  \\
\hline
DM1+QRPA & \makecell{-0.01 \\ 2.52}& \makecell{-0.01 \\ 1.11}  & \makecell{-0.01 \\ 2.52}& \makecell{-0.01 \\ 1.10}  &  \makecell{-0.04 \\ 2.88} & \makecell{-0.03 \\ 1.64} & \makecell{-0.02 \\ 2.58} & \makecell{-0.02 \\ 1.25} & \makecell{-0.01 \\ 2.55} & \makecell{0.00 \\ 1.10} & \makecell{-0.01 \\ 2.56} & \makecell{0.00 \\ 1.11} \\
\hline
SMLO  & \makecell{-0.02 \\ 2.63}& \makecell{-0.02 \\ 1.33}  & \makecell{-0.02 \\ 2.59}  & \makecell{-0.01 \\ 1.27}  &  \makecell{-0.05 \\ 3.50} &  \makecell{-0.04 \\ 2.37} & \makecell{-0.03 \\ 2.88} & \makecell{-0.03 \\ 1.66} & \makecell{-0.02 \\ 2.59}  & \makecell{-0.01 \\ 1.23} & \makecell{-0.02 \\ 2.60} &  \makecell{-0.01 \\ 1.25} \\
\hline
\end{tabular}
\end{threeparttable}
 }
\end{table}

\bibliography{supplement}